\documentclass[%
 aip,
 amsmath,amssymb,
reprint,%
]{revtex4-1}

\usepackage{graphicx}
\usepackage{dcolumn}
\usepackage{bm}
\usepackage[utf8]{inputenc}
\usepackage[T1]{fontenc}
\usepackage{mathptmx}
\usepackage{hyperref}
\begin{document}

\preprint{AIP/123-QED}

\title[Elastic Response of Wire Frame Glasses. II. 3D Systems]{Elastic Response of Wire Frame Glasses. II. Three Dimensional Systems}

\author{David A. King}
 \email{dak43@cam.ac.uk}
 \affiliation{Cavendish Laboratory, University of Cambridge, J. J. Thomson Ave., Cambridge CB3 0HE, UK}
\author{Masao Doi}%
\affiliation{Centre of Soft Matter and its Applications, Beihang University, Beijing 100191, China
}%
\author{Erika Eiser}
\affiliation{Cavendish Laboratory, University of Cambridge, J. J. Thomson Ave., Cambridge CB3 0HE, UK}%

\date{\today}

\begin{abstract}
We study the elastic response of rigid, wire frame particles in concentrated, glassy suspensions to a step strain by applying the simple, geometric methods developed in part I. The wire frame particles are comprised of thin, rigid rods of length $L$ and their number density, $\rho$, is such that $\rho L^3 \gg 1$. We specifically compare rigid rods to L-shapes made of two equal length rods joined at right angles. The behaviour of wire frames is found to be strikingly different from that of rods. The linear elasticity scales like $\rho^3 L^6$ for L-shaped particles, whereas it scales proportional to $\rho$ for rods and the non-linear response shows a transition from shear hardening to shear softening at a critical density $\rho_c \sim \sqrt{K / k_B T L^6}$, where $K$ is the bending modulus of the particles. For realistic particles made of double stranded DNA, this transition occurs at densities of about $\rho L^3 \sim 10$. The reason for these differences is that wire frames can be forced to bend by the entanglements with their surroundings, whereas rods always remain straight. This is found to be very important even for small strains, with most particles being bent above a critical strain $\gamma_c \sim (\rho L^3)^{-1}$. 
\end{abstract}

\maketitle
\section{Introduction}
\label{sec:Intro}

In the previous paper (part I)\cite{King2021ElasticModel}, we introduced a simple geometric method for determining the initial stress response to a step strain of a dense, glassy suspension of wire frame particles. In part I, the method was applied to a two dimensional model system which could be treated easily. That model showed that there was a significant difference between straight rod-like particles and bent or branched wire frames in both the magnitude of the linear elastic response and character of the non-linear behaviour. Here, we extend our discussion to three dimensional systems.

\begin{figure*}\includegraphics[width=17.2cm]{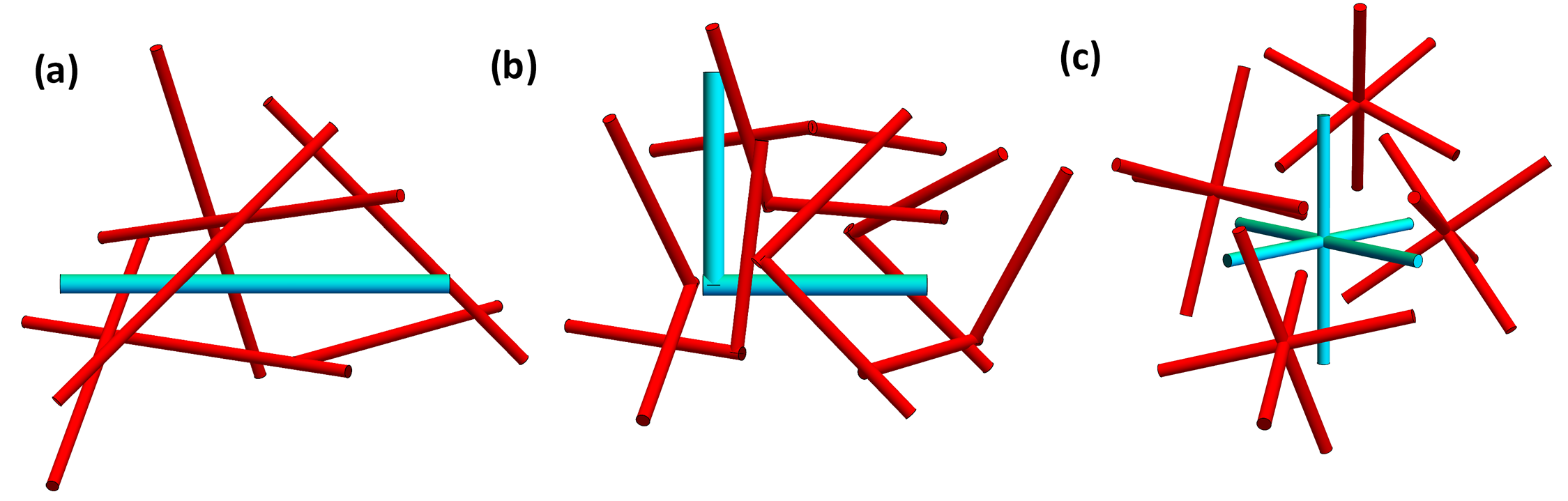}
	\caption{\label{fig:systemsketch}Sketches of the systems considered. (a) A dense suspension of rod like particles. A test rod is shown in cyan. The motion of the test rod is restricted by the surrounding red rods, but it can still diffuse along its length, so the suspension is still in a fluid state. Panels (b) and (c) show examples of dense suspensions of wire frame particles. (b) shows a test L-shaped particle in cyan and a 3D cross particle is shown in (c). These particles are completely trapped by their surroundings, shown in red. If they moves along the length of one of their legs, the other becomes entangled. The system is frozen in a glassy state.}
\end{figure*}

The wire frame particles are comprised of connected, thin, rigid rods of length $L$. The joints between the rods are taken to be effectively rigid. We are specifically interested in comparing suspensions of straight rods, as in Fig.(\ref{fig:systemsketch}a), to suspensions of bent and branched wire frames, such as L-shapes and 3D Crosses shown in Figs.(\ref{fig:systemsketch}b\&c) respectively. For simplicity, in this paper we will often only consider simple wire frame shapes, such as L-shapes, but the general conclusions will apply to other, more complicated, bent and branched shapes.

We consider number densities, $\rho$, in the range $1/L^3 \ll \rho \ll 1/V_{exc}$, where $V_{exc}$ is the extremely small excluded volume of the particles. This means the most important interactions between the particles are their topological entanglements, and there is no long range orientational order. In this concentration range, the entanglements lead to glassy dynamical behaviour for the wire frames\cite{VanKetel2005StructuralGas}, where the particles are effectively immobile, constrained to only a small set of positions and orientations at any given moment. This is not the case for rods which can still diffuse via the reptation mechanism\cite{Doi1975RotationalSolution, Doi1986TheDynamics}. This crucial difference is expected to lead to markedly different flow behaviour, which has been observed in simulations \cite{Heine2010EffectSuspensions, Petersen2010ShearNanoparticles}. The striking differences in the 2D model of part I are also expected to carry over to 3D. Understanding and predicting the differences in behaviour of suspensions of these particles has relevance for the design of DNA nanostars as functional materials \cite{Bomboi2019Cold-swappableGels,Biffi2013PhaseNanostars,Xing2018MicrorheologyHydrogels}. It is also an interesting theoretical problem since the well known methods for finding the stress in concentrated suspensions of rods based on the tube model\cite{Doi1986TheDynamics} may not be used, as they rely on the rods being able to diffuse via reptation. As such, a new method is required which determines the stress directly from the the kinetic constraints placed on the wire frame particles. We briefly review the general methods introduced in part I before applying them in three dimensions. 

Our goal is to calculate the initial elastic response of a dense suspension of wire frame particles in a glassy state to a step, shear strain. We do this by considering the change in free energy, $F$, under shear deformation. This is related to the shear elastic stress, $\sigma$, by the virtual work principle,
\begin{equation}
    \sigma = \frac{\partial F}{\partial \gamma}, 
\end{equation}
where $\gamma$ is the shear strain.

In the glassy state, the translational diffusion coefficient for the wire frames goes to zero. Therefore we treat their centres of mass as fixed. We denote the orientation of a test particle by $\Gamma$, which in three dimensions may be represented by three Euler angles. The entanglements between the particles mean that each one may only access a small region of the configuration space. The volume of this region is for a given test particle in orientation $\Gamma$ is $\Omega(\Gamma, C)$, where $C$ specifies the configuration of the constraints. The entropy associated with the test particle follows from the Boltzmann definition, 
\begin{equation}
    S(\Gamma, C) = k_B T \log \Omega(\Gamma, C). 
\end{equation}
When the step strain is applied, all of the surrounding particles move, and the volume of configuration space accessible to the test particle changes to $\tilde{\Omega}(\gamma)$. This will change the entropy, and therefore the free energy, which leads to the elastic stress. A similar method was introduced by Edwards to study flexible polymers with topological constraints and crosslinks\cite{Edwards1967StatisticalI,Edwards1969TheoryMaterial,Edwards_1969}. As the surrounding particles relax to equilibrium after the application of the strain, the test particle can explore more configurations and the stress in turn relaxes. We are interested in the magnitude of the elastic stress immediately after the deformation. Therefore we only need to consider the instantaneous change in the free energy. 

If there are $\rho$ particles per unit volume, the contribution to the free energy per unit volume from the instantaneous change in entropy is, 
\begin{equation}
\label{Fenergy1}
F_S(\gamma) = - k_B T \rho \int d\Gamma \psi(\Gamma) \Big\langle\log \tilde{\Omega}(\Gamma, C ; \gamma)\Big\rangle,
\end{equation}
where $\psi(\Gamma)$ is the orientational distribution function, which may be taken to be uniform since the system in a disordered state, and $\langle \cdots \rangle$ denotes averaging over all realisations of the surrounding constraints, $C$. 

Equation (\ref{Fenergy1}) is only the entropic contribution to the free energy and does not include any change to the internal energy when the shear is applied. As was shown in part I, this is only valid for the linear elasticity and the change in internal energy must be considered for the non-linear response. We consider the system in the absence of an external potential, therefore the internal energy only changes if the particles themselves are bent. In part I, we showed that this is only possible for wire frames, not rods. This bending mechanism must be included to understand the non-linear elasticity of wire frame glasses. 

To account for the possibility of bending, we introduce the function, $P(\gamma ; \Gamma)$. This is the probability that a particle in configuration $\Gamma$ has \textit{not} bent at strain $\gamma$. In part I we argued from simple principles that this function should be of the form
\begin{equation}
\label{Pgamma}
    P(\gamma ; \Gamma) = 1 - \frac{\gamma^2}{\gamma_{c}^2(\Gamma)},
\end{equation}
for small $\gamma$. The quantity $\gamma_c(\Gamma)$ should be interpreted as the critical strain above which all particles in configuration $\Gamma$ have bent. The bending contribution to the free energy can be determined using (\ref{Pgamma}). 

At an applied strain of $\gamma$, a particle in orientation $\Gamma$ contributes bending energy if it \textit{first} bent at a strain $\gamma' < \gamma$. The probability of this happening is, $- \partial P / \partial \gamma$ evaluated at $\gamma'$. In this case, the particle will have bent through an angle $\vartheta = (\gamma - \gamma') \Delta(\Gamma)$, where $\Delta$ depends on the particle geometry. The elastic energy associated with this bending is $K \vartheta^2/2$, where $K$ is the bending modulus of the particle. The bending contribution to the free energy can then be determined by averaging over the orientations of the particle and the strains at which it first bends,
\begin{equation}
    F_B(\gamma) = -\frac{K}{2}\rho \int d\Gamma\psi(\Gamma) \Delta^2(\Gamma) \int_{0}^{\gamma} d\gamma' (\gamma - \gamma')^2 \frac{\partial P(\gamma;\Gamma)}{\partial \gamma}\bigg\lvert_{\gamma'}.
\end{equation}
Using the general from (\ref{Pgamma}), the $\gamma'$ integral can be taken to give,
\begin{equation}
\label{bendFreeEn}
    F_B(\gamma) = \frac{K}{12}\rho \gamma^4 \int d\Gamma\psi(\Gamma)\frac{\Delta^2(\Gamma)}{\gamma_c^2(\Gamma)}.
\end{equation}
As this is proportional to $\gamma^4$, this contribution is only relevant for the non-linear elasticity. The total free energy including both the entropic and the bending contributions, valid to $\mathcal{O}(\gamma^4)$ then follows,
\begin{equation}
\label{Fenergy}
\begin{split}
F(\gamma) =& - k_B T \rho \int d\Gamma \psi(\Gamma) P(\gamma; \Gamma) \Big\langle\log \tilde{\Omega}(\Gamma ; \gamma)\Big\rangle \\
&+\frac{K}{12}\rho \gamma^4 \int d\Gamma\psi(\Gamma)\frac{\Delta^2(\Gamma)}{\gamma_c^2(\Gamma)}. \end{split}
\end{equation}
It is important to note that the first term in (\ref{Fenergy}) now must contain a factor of $P$ compared with (\ref{Fenergy1}). This is because when the particle bends, its orientation is completely determined by the constraints and therefore cannot contribute entropically.

With equation (\ref{Fenergy}) for the free energy we may determine the elastic stress for the wire frame glass from purely geometric considerations. In three dimensions it is difficult to make progress exactly without resorting to heavy algebra. In this paper we aim to explain how the key features of the 2D model found in part I carry over to 3D, using simple arguments based on the understanding gained from part I. We construct a simple model of the 3D systems in the next section. This is used to understand the bending mechanism in section \ref{sec:probbend}, which is the crucial difference between wire frame particles and straight rods. This understanding is used to compute the linear elastic response for rods and wire frames in section \ref{sec:linear}. The consistency of the model is tested by comparing the result for rigid rods to the well known Doi \& Edwards theory \cite{Doi1986TheDynamics} and we find exact agreement to leading order in density. The non-linear elastic response is considered in section \ref{sec:nonlinear} before our results are discussed in reference to real DNA nanostar systems in section \ref{sec:discussion}. 
\section{3D Model}
\label{sec:3DModel}
\begin{figure}\includegraphics[width=8cm]{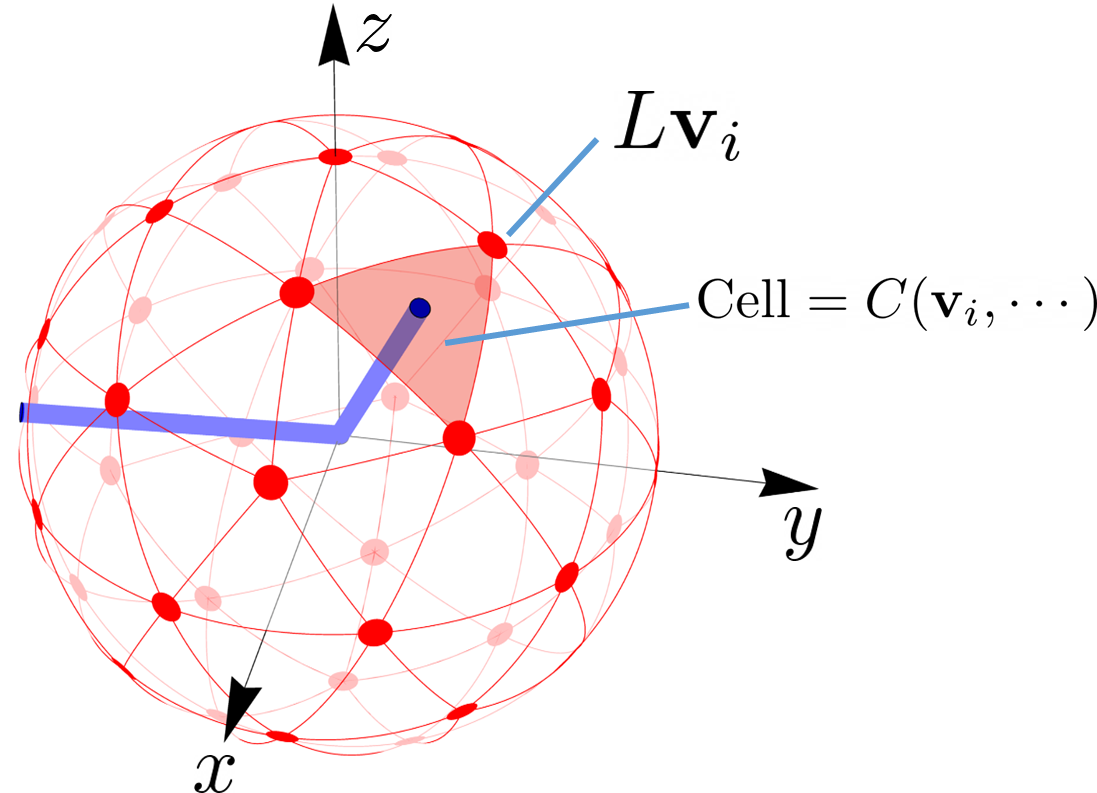}
	\caption{\label{fig:ConstCells} A sketch of the 3D model we consider. The intersections of legs of surrounding particles with the sphere encircling the blue, L-shaped, test particle are shown as red points on that sphere. The position vectors of these points are, $L \textbf{v}_i$. These points are connected together by red lines to make triangular cells which tessellate the surface of the sphere. The legs of the test particle are constrained to lie in one of these cells. The cell occupied by one leg of the test particle in this case is shown shaded in red. The identification of the cells occupied by the particle totally determines its orientation as well as $\Omega$, and all properties of the cells are determined by the vectors of its vertices, $\textbf{v}_i$.}
\end{figure}

The model we will use for these 3D systems is constructed by analogy to that used in part I for 2D systems. We consider a single test particle whose centre is fixed at the origin. The rotations of the particle about its centre are restricted by legs of surrounding particles which pierce a sphere of radius $L$ centred at the origin, surrounding the test particle. We represent these intersections by points on the surface of the sphere, the position vectors of which are given by $L \textbf{v}_i$, where $\textbf{v}_i$ is a unit vector. The configuration of the constraints is determined by the set of these unit vectors, $C = C(\textbf{v}_1, \textbf{v}_2, \cdots)$. The intersection points may be connected to each other to form triangular ``cells'' tessellating the surface of the sphere. Each leg of the test particle is then constrained to lie within one of these cells. This is sketched in Fig.(\ref{fig:ConstCells}), with an L-shaped, test particle shown in blue, the intersection points shown in red joined by red lines to form the cells. The cell occupied by one leg of the test particle is shown shaded in red. The unit vectors $\textbf{v}_i$ determine the vertices of the cells and can be used to find the accessible volume of configuration space, which is written, $\Omega(\Gamma; C(\{\textbf{v}_i\}))$. The orientations of each leg of the particle are identified as the centres of the cells which they respectively occupy, which determines $\Gamma$. 

When the shear transformation is applied, we assume that each intersection point is moved affinely so that the unit vectors $\textbf{v}_i$ transform according to, 
\begin{equation}
\label{vtrans}
    \textbf{v}_i \to \tilde{\textbf{v}}_i = \frac{(\mathbb{I}+ \kappa) \cdot \textbf{v}_i}{|(\mathbb{I}+ \kappa) \cdot \textbf{v}_i|},
\end{equation}
where $\kappa$ is the strain tensor. Each cell then changes shape accordingly and the accessible volume of configuration space transforms;
\begin{equation}
    \Omega \big(\Gamma ; C(\{\textbf{v}_i\})\big) \to \tilde{\Omega}\big(\Gamma ; \tilde{C}(\{\tilde{\textbf{v}}_i\})\big) 
\end{equation}
This formulation of the problem can be used for detailed calculations but it also allows for the important physical features to be determined from simple considerations.
\section{Bending Probability}
\label{sec:probbend}
In part I it was shown that the most important difference between straight, rod-like particles and general wire frames was that the wire frames can be forced to bend by the surroundings when the shear stress is applied, whereas rods always remain straight. This is quantified by the function $P(\gamma ; \Gamma)$, as introduced in (\ref{Pgamma}). This is always one for rigid rods, because the transformation (\ref{vtrans}) is one to one. This means all the vertices of each cell tessellating the surface of the sphere are transformed to distinct points, and no cell vanishes or overlaps with another. The value of $\Omega$ for a rod is always given by the surface area of one of these cells, and is therefore never zero or negative; bending is impossible. We wish to estimate the critical strain $\gamma_c$ above which a general wire frame particle will begin to bend. 
\begin{figure}\includegraphics[width=8cm]{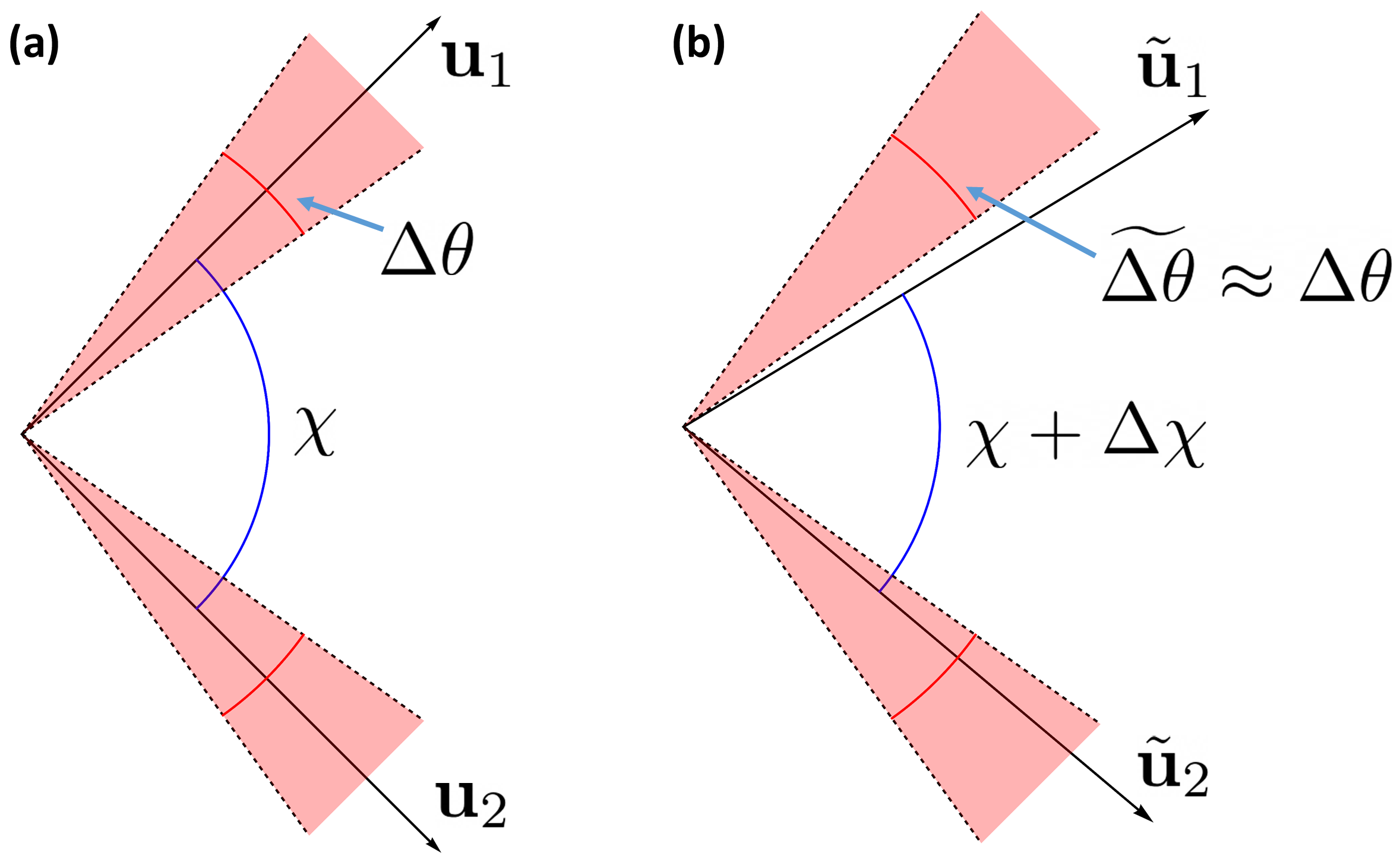}
	\caption{\label{fig:GammaCrit}(a) A sketch of configuration of a kinked, test particle with opening angle $\chi$. Shown before the step shear is applied and as a cross section in the plane of the particle. Both legs of the particle are constrained to lie in the shaded red cells which each have angular size $\Delta \theta$. The central vectors of each cell are $\textbf{u}_1$ and $\textbf{u}_2$ which make an angle of $\chi$ to each other. (b) After the strain is applied, the central vectors are transformed to $\tilde{\textbf{u}}_1$ and $\tilde{\textbf{u}}_2$ and the angle between them is now $\chi + \Delta \chi$. The average angular size of a cell after the shear is the same as before, $\widetilde{\Delta \theta} \approx \Delta \theta$. If the change $\Delta \chi$ is bigger than $\Delta \theta$ on average, then the test particle cannot maintain its original shape and it must bend.}
\end{figure}

To make this estimation we may consider a simple particle shape. A particularly illustrative choice is a `kinked' particle, a generalisation of an L-shape where the two equal length legs are joined at an angle $\chi$. This kind of particle was studied in detail in part I. Before the strain is applied, each leg lies in a separate cell. Let us define the two vectors pointing to the centre of each cell as $\textbf{u}_1$ and $\textbf{u}_2$. Initially these make an angle of $\chi$ to each other,
\begin{equation}
     \cos \chi = \textbf{u}_1 \cdot \textbf{u}_2.
\end{equation}
This situation before the shear is sketched as a cross section in the plane of the test particle in Fig.(\ref{fig:GammaCrit}a).

When the shear transformation is applied, the two unit vectors change according to the rule (\ref{vtrans}). This changes the angle between them to $\chi + \Delta \chi$, defined by,
\begin{equation}
\label{cosdeltachi}
    \cos (\chi + \Delta \chi) = \tilde{\textbf{u}}_1 \cdot \tilde{\textbf{u}}_2.
\end{equation}
The change in the angle will be a function of the applied strain, $\Delta \chi(\gamma)$, as well as the orientation of the particle which is described by the two unit vectors $\textbf{u}_{1}$ and $\textbf{u}_2$. If, on average, the size of this change in angle is larger than the average angular size of a cell, $\Delta \theta$, then the particle cannot maintain its original shape. This is shown in Fig.(\ref{fig:GammaCrit}b). Let us define the average size of $\Delta \chi$ as its root mean square value, 
\begin{equation}
\label{rmschi}
    \overline{\Delta \chi}(\gamma) = \sqrt{\langle \Delta \chi^2(\gamma) \rangle},
\end{equation}
the average here is taken over an isotropic distribution for the vectors $\textbf{u}_1$ and $\textbf{u}_2$. The particle will be forced to bend if $\overline{\Delta \chi} > \Delta\theta$, hence the critical strain for bending is estimated from, 
\begin{equation}
\label{gammaccondition}
   \overline{\Delta \chi}(\gamma_c) \sim \Delta \theta. 
\end{equation}
The average squared change in angle is found by assuming $\Delta \chi \sim \gamma$, expanding both sides of (\ref{cosdeltachi}) and matching them to first order in the strain so that,  
\begin{equation}
\label{deltachi}
 \Delta \chi \sin \chi =  \kappa_{\alpha \beta} (u_1^{\alpha}u_1^{\beta}+u_2^{\alpha}u_2^{\beta}) \cos \chi - (\kappa_{\alpha \beta} + \kappa_{\beta \alpha}) u_1^{\alpha} u_2^{\beta},
\end{equation}
here and henceforth summation is implied over repeated indices. From which it may be shown that (see the appendix\ref{app:MSCHI} for details),
\begin{equation}
\label{mschi}
 \langle\Delta \chi^2(\gamma)\rangle = \frac{1}{5}\gamma^2 \sin^2 \chi.
\end{equation}
The average angular size of a cell will be approximately the same both before and after the shear transformation. This can be estimated straightforwardly by using an argument similar to that used to find the tube radius for the reptation of rods\cite{Doi1975RotationalSolution,Doi1986TheDynamics}. We define $\Delta \theta$ as the average angle through which the particle must be rotated to \textit{first} come into contact with a surrounding particle. When the particle is rotated about any axis each of its legs sweep out a plane. The constraints on the test particle are imposed by particles intersecting this plane. The area, $a$, of the region swept out by rotating the particle by $\Delta \theta$ is approximately, 
\begin{equation}
 a \sim L^2 \Delta \theta.   
\end{equation}
Therefore, the average number of particles intersecting this region can be approximated as, 
\begin{equation}
    N \sim \rho L a \sim \rho L^3 \Delta \theta.
\end{equation}
Choosing $\Delta \theta$ so that $N \sim 1$, determines the average angular size of each cell, 
\begin{equation}
    \Delta\theta \sim \frac{1}{\rho L^3}. 
\end{equation}
Then using (\ref{rmschi}), (\ref{mschi}) and the condition (\ref{gammaccondition}) for the critical strain, we have
\begin{equation}
\label{gammac}
    \gamma_c \sim \frac{1}{\rho L^3 |\sin \chi|}.
\end{equation}
This is a very illuminating result, showing explicitly the difference between straight rods and bent wire frame particles. When the rod is straight, $\chi \to 0,\pi$, this critical strain diverges. This shows that rods never bend, no matter the applied strain. On the other hand, if the particle is even slightly kinked, then the critical strain decreases to an extremely small value $\sim (\rho L^3)^{-1}$. We can estimate for which particle shapes the bending mechanism becomes important by finding the opening angle $\chi_c$ where $\gamma_c \sim 1$. In the concentrated limit, this translates to,
\begin{equation}
    \chi_c \sim \frac{1}{\rho L^3}. 
\end{equation}
So for particles with opening angle $\chi \gtrsim \chi_c$, L-shapes or 3D crosses for example, the suspensions behaviour will be dominated by the bending mechanism. In the next sections it will be shown that this is responsible for markedly different behaviour of suspensions of wire frame particles compared to suspensions of rods. 
\section{Linear Response}
\label{sec:linear}
Initially we focus on the linear response of these systems. This requires computing the entropic free energy as in (\ref{Fenergy1}). The additional factor of $P(\gamma ; \Gamma)$ included in equation (\ref{Fenergy}) is not needed here as it only effects the non-linear stress. To check the accuracy of the model we have constructed, we compare its results for rods to the well known Doi \& Edwards theory\cite{Doi1986TheDynamics}, which is based on the tube model. We then discuss the linear elasticity for wire frame particles, by using the simple example of an L-shaped particle. 
\subsection{Rods}
\label{sec:Rods}
For rod-like particles, the model can be used to calculate the linear elastic stress exactly, to leading order in the density. The orientation of the rod is specified by the unit vector $\textbf{u}$ running parallel to its length. The constraints on the particle are defined by the cell, $C$, which it occupies. This cell is defined by the three vectors of its vertices, $\textbf{v}_1$, $\textbf{v}_2$ and $\textbf{v}_3$. The accessible volume of configuration space is written, $\Omega(\textbf{u} ; C)$. When the shear transformation is applied, the vertices of $C$ move according to the transformation rule (\ref{vtrans}) and the accessible volume changes to $\tilde{\Omega}(\textbf{u}; \tilde{C})$. 
\begin{figure}\includegraphics[width=8cm]{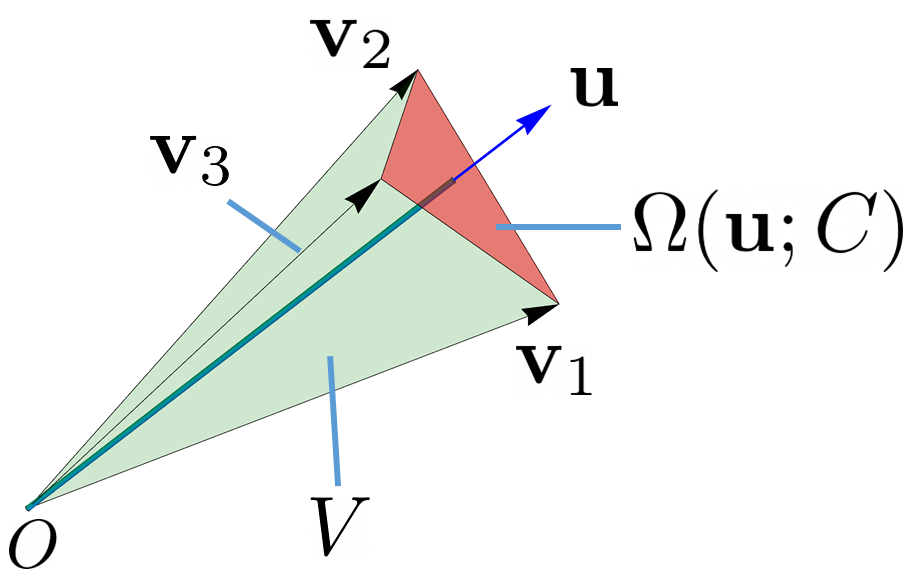}
	\caption{\label{fig:Polyhedron} A sketch of a test rod, shown in blue, with orientation parallel to $\textbf{u}$ which is constrained by the surroundings to lie in the red cell whose area is $\Omega(\textbf{u};C)$. The vertices of the cell are at positions given by $\textbf{v}_1$, $\textbf{v}_2$ and $\textbf{v}_3$. The area $\Omega$ can be determined by first finding the volume, $V$ of the green polyhedron whose vertices are the origin and those of the red cell.}
\end{figure}

To determine the entropy, we must compute the ratio $\tilde{\Omega}/\Omega$. This can be calculated by considering the change in volume of the polyhedron whose vertices are at the origin, $\textbf{v}_1$, $\textbf{v}_2$ and $\textbf{v}_3$. The polyhedron is sketched in Fig.(\ref{fig:Polyhedron}). Using simple geometry, the volume of this polyhedron before the shear is, 
\begin{equation}
\label{Vol1}
    V = \int d\Omega \int_0^{|\textbf{u}|} dr \ r^2 \equiv \frac{1}{3} | \textbf{u} |^3 \Omega(\textbf{u};C).
\end{equation}
After the shear is applied the volume changes by definition to, 
\begin{equation}
\label{Vol2}
    \tilde{V} = \det(\mathbb{I} + \kappa) V.
\end{equation}
This can also be written as, 
\begin{equation}
\label{Vol3}
    \tilde{V} = \frac{1}{3} |(\mathbb{I} + \kappa) \cdot \textbf{u} |^3 \tilde{\Omega}(\textbf{u};\tilde{C}).
\end{equation}
Dividing (\ref{Vol3}) by (\ref{Vol1}) and using (\ref{Vol2}) yields the ratio,
\begin{equation}
\frac{\tilde{\Omega}(\textbf{u};\tilde{C})}{\Omega(\textbf{u};C)} = \frac{|\textbf{u}|^3}{ |(\mathbb{I} + \kappa) \cdot \textbf{u}|^3} \det(\mathbb{I}+\kappa).
\end{equation}
To calculate the entropy, we need to take the logarithm of this ratio,
\begin{equation}
\label{logOmegarod}
\log \frac{\tilde{\Omega}(\textbf{u};\tilde{C})}{\Omega(\textbf{u};C)} = 3 \log |\textbf{u}| - 3 \log |(\mathbb{I} + \kappa) \cdot \textbf{u}| + \log \det(\mathbb{I}+\kappa).
\end{equation}
The final term of this can be treated using the identity, $\log \det A = \text{Tr}\log A$, which holds for any square, non-singular matrix $A$. This yields
\begin{equation}
\text{Tr}\log(\mathbb{I} + \kappa) =  - \frac{1}{2}\text{Tr}\kappa^2 + \mathcal{O}(\kappa^3).    
\end{equation}
The first order term in $\kappa$ in this expansion vanishes because of the incompressiblilty condition, $\text{Tr}\kappa =0$.

The other two terms in (\ref{logOmegarod}) are now expanded to second order in $\kappa$,
\begin{equation}
\log \tilde{\Omega}/\Omega = - 3 \textbf{u} \cdot \kappa \cdot \textbf{u} - \frac{3}{2} \big[|\kappa \cdot \textbf{u}|^2 - 2 (\textbf{u} \cdot \kappa \cdot \textbf{u})^2\big] - \frac{1}{2} \text{Tr}\kappa^2.
\end{equation}
The entropic free energy can now be obtained by averaging over $\textbf{u}$, which is written in components as,
\begin{equation}
\begin{split}
    \frac{F_s(\kappa)}{k_B T} &= 3\rho \kappa_{\alpha \beta} \langle u_{\alpha} u_{\beta} \rangle + \frac{3}{2}\rho \kappa_{\mu \alpha} \kappa_{\mu \beta} \langle u_{\alpha} u_{\beta} \rangle\\
    &- 3\rho  \kappa_{\alpha \beta}\kappa_{\mu \nu} \langle u_{\alpha}u_{\beta}u_{\mu}u_{\nu} \rangle - \frac{1}{2}\rho \kappa_{\alpha \beta} \kappa_{\beta \alpha}. 
\end{split}
\end{equation}
To lowest order in density, the distribution of $\textbf{u}$ can be taken to be isotropic. In which case the averages are,
\begin{subequations}
\begin{equation}
    \langle u_{\alpha} u_{\beta} \rangle = \frac{1}{3} \delta_{\alpha \beta},
\end{equation}
and
\begin{equation}
    \langle u_{\alpha}u_{\beta}u_{\mu}u_{\nu} \rangle = \frac{1}{15}(\delta_{\alpha \beta} \delta_{\mu \nu} + \delta_{\alpha \mu} \delta_{\beta \nu} + \delta_{\alpha \nu} \delta_{\beta \mu}).
\end{equation}
\end{subequations}
So that the free energy becomes,
\begin{equation}
    \frac{F(\kappa)}{k_B T} = \frac{3}{10}\rho \kappa_{\alpha \beta} \kappa_{\alpha \beta}  - \frac{6}{5}\rho \kappa_{\alpha \beta} \kappa_{\beta \alpha}. 
\end{equation}
Where once again we have used the fact that $\kappa$ is traceless. We now restrict our attention to the case of simple shear, where $\kappa$ has only one non-vanishing component $\kappa_{xy}= \gamma$.  In this case, it is straightforward to show that, 
\begin{equation}
    F(\kappa) = \frac{3}{10} k_B T \rho \gamma^2,
\end{equation}
and the linear elastic stress follows immediately,
\begin{equation}
\label{stressrod}
    \sigma = \frac{3}{5} k_B T \rho \gamma.
\end{equation}
This result is in precise agreement with the well known Doi \& Edwards result. This demonstrates the consistency of our simple 3D model. 
\subsection{Wire Frames}
\label{sec:Lshapes}
An explicit calculation of the stress for general wire frame particles in 3D is very complicated. However, the understanding developed in part I can be used to determine the scaling of the stress from a simple argument. As discussed in section \ref{sec:probbend}, the crucial difference between general wire frame particles and rods is the possibility of bending. While rods can keep their shape for arbitrary $\gamma$, wire frame particles cannot and begin to bend above a certain strain. We will exploit this fact to estimate $\Omega$ for general wire frames.

Let us imagine a perfectly flexible L-shaped wire frame particle. The legs of such a particle are freely jointed so that the angle between the two legs can take any value without incurring any energetic cost. No matter the size of the applied strain, this particle can always access a range of configurations. The accessible volume of configuration space for this flexible particle after the strain is applied is defined to be, $\tilde{\Omega}_{\text{flex}}(\Gamma ; \tilde{C})$. Up until the strain where the rigid particle begins to bend, it may access the same volume of configuration space as the imagined flexible particle. However, above $\gamma_c$, the rigid wire frame particle bends and $\tilde{\Omega}_{\text{rigid}} = 0$. Hence we may write, 
\begin{equation}
\label{rigidomega}
\tilde{\Omega}_{\text{rigid}}(\Gamma ; \tilde{C}) = P(\gamma ; \Gamma) \tilde{\Omega}_{\text{flex}}(\Gamma ; \tilde{C}). 
\end{equation}
Substituting this into the expression for the free energy (\ref{Fenergy1}) gives, 
\begin{equation}
\label{FreeEnergyWireFrame1}
    \frac{F_s(\gamma)}{k_B T} = - \rho \big\langle \log P(\gamma ; \Gamma) \big\rangle - \rho \big\langle \log \tilde{\Omega}_{\text{flex}}(\Gamma ; \tilde{C}) \big\rangle, 
\end{equation}
where the angle brackets denote averaging over orientations $\Gamma$ and constraint configurations $C$. The second term may be calculated in a similar manner as for rods in the previous section. Each leg of the L-shape is independent of the other, so the free energy associated with each is essentially the same as that of a singe rod. Therefore, this term must be of the form, 
\begin{equation}
    \rho \big\langle \log \tilde{\Omega}_{\text{flex}}(\Gamma ; \tilde{C}) \big\rangle =- c \rho  \gamma^2,
\end{equation}
where $c$ is a positive constant of order unity. The first term in (\ref{FreeEnergyWireFrame1}), however, is very different. Using (\ref{Pgamma}) we have, 
\begin{equation}
    \rho \big\langle \log P(\gamma ; \Gamma) \big\rangle = \rho \big\langle \log \big(1 - \gamma^2/\gamma_c^2(\Gamma)\big) \big\rangle.
\end{equation}
This may be expanded to $\mathcal{O}(\gamma^2)$ to obtain the free energy relevant for the linear elasticity of wire frame systems, 
\begin{equation}
    \frac{F(\gamma)}{k_B T} = \rho \bigg\langle \frac{\gamma^2}{\gamma_c^2(\Gamma)}\bigg\rangle + c \rho \gamma^2.
\end{equation}
Replacing $\gamma_c(\Gamma)$ by its average, and using its scaling found in (\ref{gammac}), this becomes, 
\begin{equation}
    \frac{F(\gamma)}{k_B T} = a \rho(\rho L^3)^2  \gamma^2 + c \rho \gamma^2,
\end{equation}
where $a$ is another constant. For the dense suspensions we consider, where $\rho L^3 \gg 1$, the first term dominates the second and the free energy scales like, 
\begin{equation}
    F(\gamma) \sim k_B T \rho(\rho L^3)^2 \gamma^2.
\end{equation}
It follows then that the linear elastic stress for concentrated suspensions of L-shapes scales as,
\begin{equation}
\label{LinearStressLshape}
    \sigma \sim k_B T \rho^3 L^6 \gamma.
\end{equation}
This is far larger than that found for rods in (\ref{stressrod}), and much more sensitive to the density and size of the particle. This is qualitatively consistent with the simulation results\cite{Heine2010EffectSuspensions,Petersen2010ShearNanoparticles} for dense suspensions of 3D crosses, sketched in Fig.(\ref{fig:systemsketch}c). Those simulations showed that the zero shear rate viscosity of a dense suspension of these particles increased by approximately three orders of magnitude when the volume fraction was increased from $0$ to $0.1$. It is difficult to quantitatively compare these results to ours because the density dependence of the zero shear rate viscosity is also determined that of the relaxation timescale of the stress, which is not addressed in this series of papers. We hope that the methods we have introduced can provide a useful framework for studying this problem and more accurate comparisons can be made in the future.    

The argument presented in this section shows that the bending mechanism is responsible for the very large magnitude of the linear elasticity in these systems. In section \ref{sec:probbend} it was shown that $\gamma_c$ is very small and the bending mechanism becomes important for kinked particles with opening angle $\chi \gtrsim (\rho L^3)^{-1}$. Therefore, the very strong density dependence of the stress found in (\ref{LinearStressLshape}) is expected for particles which are only slightly bent away from a straight rod, through an angle of about $\chi \approx 20^{\circ}$ say. It is clear also that branched particles, such as Y-shapes or 3D crosses (shown in Fig.(\ref{fig:systemsketch}c)), should exhibit the same scaling. This discussion shows that the elastic stress in concentrated suspensions of wire frame particles is extremely sensitive to the particle shape, displaying radically different behaviour compared to rods for even modestly bent particles. The example of 3D crosses is especially noteworthy, since a dilute suspension of these particles has no elastic response for fundamental symmetry reasons\cite{King2020ParticleSuspensionsb}. This means that the behaviour of suspensions of these particles will be most sensitive to concentration, exhibiting a sharp transition from a Newtonian fluid to an elastic gel.  
\section{Non-Linear Response}
\label{sec:nonlinear}
\subsection{Rods}
For rod like particles, the extension non-linear elasticity is straightforward because there is no probability of bending, $P(\Gamma;\gamma) =1$. Therefore all that needs to be done is expand the entropy to $\mathcal{O}(\gamma^4)$. The exact calculation is long winded but we can appeal to the result in part I, as well as results based on the tube model for the general form. These show that the suspension shear thins, with the stress to $\mathcal{O}(\gamma^3)$ being given by, 
\begin{equation}
\label{nonlinstressrods}
    \sigma = k_B T \rho (a \gamma - b \gamma^3),
\end{equation}
where $a$ and $b$ are positive, order unity and approximately independent of density for $\rho L^3 \gg 1$. 
\subsection{Wire Frames}
\label{sec:wireframenonlinear}
For wire frame particles, both the entropic and bending contribution to the free energy are required. In particular, for L-shaped particles, the entropic free energy can be estimated from the expression for the accessible configuration space volume for a rigid wire frame (\ref{rigidomega}) introduced in section \ref{sec:Lshapes}. From the first term in (\ref{Fenergy}) we have,
\begin{equation}
\label{FreeEnergyWireFrame2}
    \frac{F_s(\gamma)}{k_B T} = - \rho \big\langle P(\gamma ; \Gamma) \log P(\gamma ; \Gamma) \big\rangle - \rho \big\langle P(\gamma ; \Gamma) \log \tilde{\Omega}_{\text{flex}}(\Gamma ; \tilde{C}) \big\rangle. 
\end{equation}
For simplicity, we introduce the following pre-averaging approximations,
\begin{subequations}
\begin{equation}
    \big\langle P(\gamma ; \Gamma) \log P(\gamma ; \Gamma) \big\rangle \approx \big\langle P(\gamma ; \Gamma)\big\rangle \log \big\langle P(\gamma ; \Gamma) \big\rangle,
\end{equation}
and
\begin{equation}
    \big\langle P(\gamma ; \Gamma) \log \tilde{\Omega}_{\text{flex}}(\Gamma ; \tilde{C}) \big\rangle \approx \big\langle P(\gamma ; \Gamma)\big\rangle \big\langle \log \tilde{\Omega}_{\text{flex}}(\Gamma ; \tilde{C}) \big\rangle.
\end{equation}
\end{subequations}
These approximations make our analysis simpler, but do not change the nature of our results.

Due to the fact that legs of the flexible L-shape behave as independent rods, we can write
\begin{equation}
\big\langle \log \tilde{\Omega}_{\text{flex}}(\Gamma ; \tilde{C}) \big\rangle \sim -\frac{a}{2} \gamma^2 + \frac{b}{4} \gamma^4,
\end{equation}
where $a$ and $b$ are positive and approximately independent of $\rho$, so that the stress arising from this term is the same form as (\ref{nonlinstressrods}) for rods. Along with the definition (\ref{Pgamma}), this allows us to write the entropic free energy as, 
\begin{equation}
\label{FreeEnergyWireFrame3}
    \frac{F_s(\gamma)}{k_B T} = - \rho \bigg(1 - \frac{\gamma^2}{\gamma_c^2}\bigg)\bigg[\log\bigg(1 - \frac{\gamma^2}{\gamma_c^2}\bigg) - \frac{a}{2} \gamma^2 + \frac{b}{4} \gamma^4\Bigg].
\end{equation}
Which, when expanded to fourth order in $\gamma$ gives
\begin{equation}
\label{FreeEnergyWireFrame4}
    \frac{F_s(\gamma)}{k_B T} = \rho \bigg(\frac{1}{\gamma_c^2} + \frac{a}{2}\bigg)\gamma^2 - \rho \bigg(\frac{1}{2\gamma_c^4} + \frac{a}{2 \gamma_c^2} + \frac{b}{4}\bigg)\gamma^4.
\end{equation}
Using the scaling of $\gamma_c$ with density, and dropping the sub-dominant terms we obtain the general scaling form of the entropic free energy for wire frame particles,
\begin{equation}
\label{FreeEnergyWireFrame5}
    \frac{F_s(\gamma)}{k_B T} \sim \frac{g}{2} \rho (\rho L^3)^2 \gamma^2 - \frac{h}{4} \rho (\rho L^3)^4\gamma^4,
\end{equation}
where $g$ and $h$ are positive constants which depend on the particle geometry. Note that in deriving this form, we have assumed that the function $P(\gamma ; \Gamma)$ is \textit{exactly} as given in (\ref{Pgamma}) with no corrections $\mathcal{O}(\gamma^4)$. This clearly need not be the case, however including these terms does not alter the general from of (\ref{FreeEnergyWireFrame5}). The term proportional to $\gamma^4$ in the entropic free energy (\ref{FreeEnergyWireFrame5}) must appear with a negative sign since at a strain $ \propto \gamma_c$, the entropic free energy is zero. This strain also approximately represents the limit of the description presented here, because above this strain other effects, such as the non-linear elasticity of the particles themselves, will begin to play a role. 

The bending free energy is calculated directly from (\ref{bendFreeEn}),
\begin{equation}
\label{bendFreeEn2}
    F_B(\gamma) \sim \frac{K}{4} \rho (\rho L^3)^2 C_B \gamma^4,
\end{equation}
where we have defined the positive constant, 
\begin{equation}
    C_B =\frac{1}{3(\rho L^3)^2}\bigg\langle \frac{\Delta^2(\Gamma)}{\gamma_c^2(\Gamma)}\bigg\rangle,
\end{equation}
which is order unity and depends on the particle geometry.

The total free energy is therefore the sum of (\ref{bendFreeEn2}) and (\ref{FreeEnergyWireFrame5}),
\begin{equation}
\label{FreeEnergyWireFrame6}
    \frac{F(\gamma)}{k_B T} \sim \frac{g}{2} \rho (\rho L^3)^2 \gamma^2 + \frac{1}{4} \rho (\rho L^3)^2 \bigg(\frac{K}{k_B T} C_B  - h (\rho L^3)^2\bigg)\gamma^4,
\end{equation}
This can lead to two different behaviours for the stress, depending on the value of $K$ relative to the density. This is due to the competition between the positive and negative contributions to  the coefficient of the $\gamma^4$ term. If the coefficient of $\gamma^4$ is positive, the suspension shear hardens. This occurs when the bending modulus satisfies,
\begin{equation}
\label{Gcondition}
    \frac{K}{k_B T} \gtrsim (\rho L^3)^{2}. 
\end{equation}
This condition is non-trivial since, for our results to be valid we require both $\rho L^3 \gg 1$, and due to the rigidity of the particles, $K \gg k_B T$. We can take, $K \sim k_B T (\rho L^3)^{p}$, for any $p \geq 1$ and our results will still hold, but only for $p > 2$ will we see shear hardening. Alternatively this can be understood as a condition on the density, and equation (\ref{Gcondition}) would imply that above a critical density,
\begin{equation}
    \label{rhocondition}
    \rho_{c} \sim \sqrt{\frac{K}{k_B T L^6}},
\end{equation}
the suspension will shear thin. The exact values of $K$ and $\rho$ where the cross over between these two behaviours occurs depends on the particle geometry through the ratio $C_B / h$. 

This situation is qualitatively exactly the same as was found in part I. This behaviour has an explanation at the level of the model presented here. When a particle starts to bend, its orientation is completely determined by the surroundings, and as such cannot contribute to the entropic free energy. This effect is captured by the $P(\gamma ; \Gamma)$ factor in the first term of equation (\ref{Fenergy}). As the applied strain is increased, more and more particles begin to bend, so fewer and fewer contribute entropically. This deficit leads to the strong shear softening behaviour of $F_{s}(\gamma)$. If, at a given strain, the bending contribution is not sufficient to make up this deficit, the total stress will be shear softening. Therefore, there is some critical value of the bending modulus which must be exceeded to see a shear hardening response. 
\section{Discussion}
\label{sec:discussion}
We have discussed the elastic response of a dense suspension of rigid rod like and L-shaped particles in three dimensions. A simple geometric method is used to calculate the entropy of the system by determining the volume of configuration space accessible to a particular particle, given the constraints placed on it by its surroundings. The change in accessible volume under the transformation associated with an applied shear leads to a change in the free energy of the system. For the L-shaped particles, it is possible that they need to bend when the system is sheared in order for them to respect the constraints placed on them by their surroundings. This bending mechanism contributes to the free energy of the system. This is taken into account by introducing the function, $P(\gamma)$, interpreted as the proportion of particles which have not bent at a strain $\gamma$. We find that, to lowest order in $\gamma$, this is given by $P = 1 - (\gamma/\gamma_c)^2$, where $\gamma_c$ is the critical strain above which most particles have bent. We determined that $\gamma_c \sim (\rho L^3)^{-1}$, and found the elastic stress up to $\mathcal{O}(\gamma^3)$, valid for strains less than this critical value.

As for the 2D model presented in part I, we find two very interesting results. First, the elastic stress in a system of L-shaped particles is significantly higher than that for a system of rods, scaling proportional to $\rho^3 L^6$ as opposed to $\rho$. Since we focus on the concentrated regime with $\rho L^3 \gg 1$, this represents a drastic change in the suspensions behaviour. We also find that this behaviour is very sensitive to the shape of the suspended particles, with the new scaling present for particles bent through any angle $\sim \mathcal{O}(1)$. This conclusion also holds for any branched particle shape, e.g. 3D Crosses (see Fig.(\ref{fig:systemsketch}c)). 

Second, we find that there is a critical density above which the solution is shear softening. This density depends on the bending modulus of the particle and is approximately, $\rho_c \sim (K /k_B T)^{1/2} L^{-3}$. Conversely, if the particles are rigid enough, then the suspension shear hardens. This is in contrast to the behaviour of a rigid rod system, which always shear thins. 

A potential realisation of the kind of system discussed here are DNA nano-stars, where double stranded DNA legs are joined together at prescribed angles. Taking the length of the legs to be on the order of ten base pairs ($\sim 1$nm), we can estimate the bending modulus to be, $K \sim 50 k_B T$, which means that at concentrations $\rho L^3 \gtrsim 10$, we expect shear softening. The analysis presented here is valid in the concentration range, $1 \ll \rho L^3 \ll 50$, where the upper limit is set by the concentration at which excluded volume effects become important. We therefore hope that the transition between the two behaviours can be verified experimentally.
\acknowledgements 
We are grateful to Prof. Daan Frenkel for a number of insightful, interesting and important discussions. D.A.K. acknowledges financial support from the UK Engineering and Physical Sciences Research Council Ph.D. Studentship award No. 1948692.
\appendix*
\section{Calculation of \texorpdfstring{$\langle\Delta \chi^2\rangle$}{TEXT}}
\label{app:MSCHI}
In this appendix we give some details of the calculation of the mean squared value of $\Delta \chi$. Our starting point is equation (\ref{deltachi}) of the main text,
\begin{equation}
    \Delta \chi \sin \chi = \kappa_{\alpha \beta} (u_1^{\alpha}u_1^{\beta}+u_2^{\alpha}u_2^{\beta}) \cos \chi - (\kappa_{\alpha \beta} + \kappa_{\beta \alpha}) u_1^{\alpha} u_2^{\beta}.
\end{equation}
Taking the square of this and averaging over $\textbf{u}_1$ and $\textbf{u}_2$ gives, 
\begin{equation}
\begin{split}
    &\langle\Delta \chi^2 \rangle\sin^2 \chi = \cos^2\chi \kappa_{\alpha \beta} \kappa_{\gamma \delta}\big\langle(u_1^{\alpha}u_1^{\beta}+u_2^{\alpha}u_2^{\beta})(u_1^{\delta}u_1^{\delta}+u_2^{\gamma}u_2^{\delta})\big\rangle \\
    &+(\kappa_{\alpha \beta} + \kappa_{\beta \alpha})(\kappa_{\gamma \delta} + \kappa_{\delta \gamma})\big\langle u_1^{\alpha} u_2^{\beta}u_1^{\gamma} u_2^{\delta}\big\rangle \\
    &-2\cos\chi \kappa_{\alpha \beta} (\kappa_{\gamma \delta} + \kappa_{\delta \gamma}) \big[\big\langle u_1^{\gamma} u_1^{\alpha}u_1^{\beta} u_2^{\delta}\big\rangle+\big\langle u_2^{\delta} u_2^{\alpha}u_2^{\beta} u_1^{\gamma}\big\rangle\big].
\end{split}
\end{equation}
Let us refer to the three terms on the right hand side of this equation as $\text{I}$,$\text{II}$ and $\text{III}$ respectively so that,
\begin{subequations}
\label{appI}
\begin{equation}
\begin{split}
    \text{I} = \cos^2\chi \kappa_{\alpha \beta} \kappa_{\gamma \delta}&\big[
    \big\langle u_1^{\alpha}u_1^{\beta}u_1^{\gamma}u_1^{\delta}\big\rangle + \big\langle u_1^{\alpha}u_1^{\beta}u_2^{\gamma}u_2^{\delta})\big\rangle \\
    &+\big\langle u_1^{\gamma}u_1^{\delta}u_2^{\alpha}u_2^{\beta}\big\rangle +  \big\langle u_2^{\alpha}u_2^{\beta}u_2^{\gamma}u_2^{\delta}\big\rangle\big],
\end{split}
\end{equation}
\begin{equation}
    \text{II} = (\kappa_{\alpha \beta} + \kappa_{\beta \alpha})(\kappa_{\gamma \delta} + \kappa_{\delta \gamma})\big\langle u_1^{\alpha} u_1^{\gamma}u_2^{\beta} u_2^{\delta}\big\rangle,
\end{equation}
\begin{equation}
    \text{III} = -2\cos\chi \kappa_{\alpha \beta} (\kappa_{\gamma \delta} + \kappa_{\delta \gamma}) \big[\big\langle u_1^{\gamma} u_1^{\alpha}u_1^{\beta} u_2^{\delta}\big\rangle+\big\langle u_2^{\delta} u_2^{\alpha}u_2^{\beta} u_1^{\gamma}\big\rangle\big].
\end{equation}
\end{subequations}
The unit vectors are distributed isotropically but with a fixed angle $\chi$ between them. The averages of the fourth rank tensors in the above equations can therefore be determined easily,
\begin{subequations}
\begin{equation}
    \big\langle u_{1}^{\alpha}u_{1}^{\beta}u_{1}^{\gamma}u_{1}^{\delta} \big\rangle = \frac{1}{15}\big(\delta_{\alpha \beta} \delta_{\gamma \delta} + \delta_{\alpha \gamma} \delta_{\beta \delta} + \delta_{\alpha \delta} \delta_{\beta \gamma}\big),
\end{equation}
\begin{equation}
    \big\langle u_{1}^{\alpha}u_{1}^{\beta}u_{1}^{\gamma}u_{2}^{\delta} \big\rangle = \frac{\cos \chi}{15}\big(\delta_{\alpha \beta} \delta_{\gamma \delta} + \delta_{\alpha \gamma} \delta_{\beta \delta} + \delta_{\alpha \delta} \delta_{\beta \gamma}\big),
\end{equation}
\begin{equation}
\begin{split}
    \big\langle u_{1}^{\alpha}u_{1}^{\beta}u_{2}^{\gamma}u_{2}^{\delta} \big\rangle = &\frac{1}{30}\big[2(2 - \cos^2\chi)\delta_{\alpha \beta} \delta_{\gamma \delta}\\
    &+ (3 \cos^2 \chi - 1)\big(\delta_{\alpha \gamma} \delta_{\beta \delta} + \delta_{\alpha \delta} \delta_{\beta \gamma}\big)\big].   
\end{split}
\end{equation}
\end{subequations}
Using these in (\ref{appI}) for the specific case of simple shear, where $\kappa$ has only one non-zero component $\kappa_{xy} = \gamma$, it is straight forward to show that,
\begin{subequations}
\label{appII}
\begin{equation}
\begin{split}
    \text{I} = \frac{2\gamma^2}{15} \cos^2\chi + \frac{\gamma^2}{15} \cos^2 \chi (3 \cos^2 \chi - 1),
\end{split}
\end{equation}
\begin{equation}
    \text{II} = \frac{2\gamma^2}{15} (2 - \cos^2 \chi) + \frac{\gamma^2}{15}(3 \cos^2 \chi - 1),
\end{equation}
\begin{equation}
    \text{III} = -\frac{8\gamma^2}{15} \cos^2\chi.
\end{equation}
\end{subequations}
Hence we have, 
\begin{equation}
\begin{split}
    \langle\Delta \chi^2(\gamma)\rangle \sin^2\chi = \frac{\gamma^2}{15}\big(3 - 6\cos^2 \chi + 3\cos^4 \chi\big) = \frac{1}{5}\gamma^2 \sin^4 \chi.
\end{split}
\end{equation}
From which, equation (\ref{mschi}) of the main text is easily recovered. 

\section*{Data Availability}
Data sharing is not applicable to this article as no new data were created or analyzed in this study.
\bibliography{references}
\end{document}